\documentclass[aps,pra,twocolumn,groupedaddress,showpacs, floatfix]{revtex4-1}
\usepackage{amsmath}
\usepackage{graphicx}
\usepackage{bm}
\usepackage[mathlines]{lineno}

\begin{document}

\title{Measurement of the hyperfine coupling constant for $nS_{1/2}$ Rydberg states of $^{85}$Rb}
\author{Andira Ramos} \email{andramos@umich.edu}
\author{Ryan Cardman}
\author{Georg Raithel}
\affiliation{Department of Physics, University of Michigan, Ann Arbor, Michigan 48105, USA}

\date{\today}

\begin{abstract}

We present measurements of the hyperfine structure splittings of $nS_{1/2}$ Rydberg states of $^{85}$Rb for $n=$ 43, 44, 45 and 46. From the splittings, the hyperfine coupling constant, $A_\mathrm{HFS}$, is determined to be 15.372(80)~GHz. This result is an order-of-magnitude improvement from previous measurements. We study and account for systematic uncertainty sources, such as unwanted electric and magnetic fields, dipolar Rydberg-Rydberg interactions, and AC shifts.
Initial evidence for hyperfine-mixed Rydberg pair states is found.
\end{abstract}

\maketitle


Hyperfine structure (HFS) splittings hold important information about the nucleus of an atom such as the values of the nuclear magnetic dipole, electric quadrupole and magnetic octuple moments~\cite{Steck85Rb}. These splittings also depend on the value of the electronic wavefunction at the location of the nucleus~\cite{Arimondo1977}. Moreover, hyperfine states are used in quantum computation with ions and neutral atoms~\cite{Ladd2010,HFSquantumCompRb2002, HFSquantumCompIons2014}. This makes the implications of knowing the HFS splittings well far-reaching.

Some quantum computation schemes with neutral atoms employ Rydberg states to perform operations~\cite{Lukin2001, SaffmanReview2010,SaffmanReview2016}. Therefore, knowing the HFS of Rydberg states is important for the successful implementation of these quantum operations. Furthermore, sensitive precision measurements~\cite{SafronovaReview2018} and studies of Rydberg-ground and Rydberg-Rydberg molecules involving $S$-states~\cite{Anderson2014, Sassmannshausen2015, Bottcher2016, Maclennan2019} also rely on precise knowledge of the Rydberg HFS splittings.

The most recent experiments for HFS splittings of $nS_{1/2}$ Rydberg states of rubidium have yielded uncertainties of about 60~kHz for $^{85}$Rb $nS_{1/2}$ states with principal quantum numbers $n\leq 33$ ~\cite{GallagherNsNp} (relative uncertainty of 8.9$\%$) and 100~kHz for $^{87}$Rb~\cite{SpreeuwHFS} (relative uncertainty of 2.3$\%$). In the present work, we perform measurements of $^{85}$Rb HFS splittings with uncertainties between 0.4 and 2~kHz (relative uncertainties of 1$\%$ and below). We use the measured splittings, $\nu_{\mathrm{HFS}}$, experimentally determined quantum defects, $\delta_s(n)$ \cite{GallagherNsNp}, and the relation
\begin{equation}
\label{eq:HFSsplitting1}
\nu_{\mathrm{HFS}}= \frac{A_{HFS}}{[n-\delta_s(n)]^3},
\end{equation}
\noindent to extract the HFS coupling constant, $A_{\mathrm{HFS}}$, with a relative uncertainty of 0.5$\%$. This is an improvement of an order of magnitude from the best measurement available to date, which has a relative uncertainty of 9.5$\%$~\cite{GallagherNsNp}.

\section{Procedure}

\begin{figure}[b]
\centering
\includegraphics[width=3.4in]{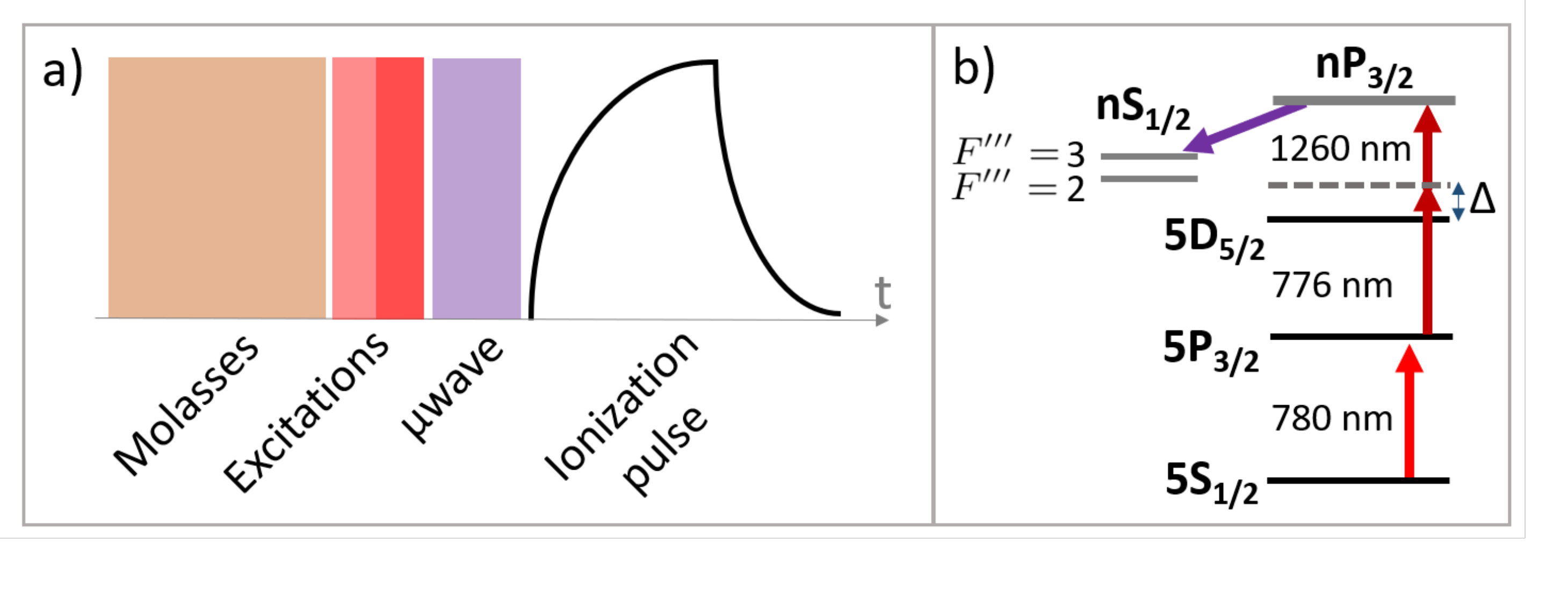}
\caption{(color online) a) Timing sequence used in each experimental cycle. b) Excitation scheme with laser wavelengths indicated; intermediate detuning $\Delta \simeq 130$~MHz. The $nP_{3/2}\rightarrow nS_{1/2}$ transition is a microwave transition. Energy splittings not to scale.}
\label{fig:HFStiming}
\end{figure}

To measure the HFS splittings, we employ microwave spectroscopy between two Rydberg states. We perform the measurements in a vacuum chamber with cold Rydberg atoms. Atoms are cooled in a bright optical molasses using the $^{85}$Rb D2 cooling and repumping transitions ~\cite{Metcalf1988, Chu1985}. To reach the Rydberg states, atoms are initially excited from $|5S_{1/2}, F=3 \rangle$ to $|5P_{3/2}, F'=4 \rangle$ (wavelength of 780~nm) followed by a two-photon transition from $|5P_{3/2}, F'=4 \rangle$ to $|nP_{3/2}, F''=2,3,4 \rangle$ using two laser beams set to $\sim$~776~nm and 1260~nm (see Fig.~\ref{fig:HFStiming}). We then scan the frequency of linearly-polarized microwave radiation to drive the transition $nP_{3/2} \rightarrow nS_{1/2}$ for $n=$ 43, 44, 45 and 46. We use an Agilent signal generator (N5183A) as a microwave source and reference it to a rubidium atomic clock (SRS model FS725). The frequency of the microwave signal is up-converted using an active frequency multiplier, the output of which is transmitted through a horn antenna into the vacuum chamber. The 780-nm excitation beam is pulsed on for 40~$\mu$s and the 776 and 1260-nm excitation beams are pulsed on for 20~$\mu$s, followed by a 40~$\mu$s microwave pulse (see Fig.~\ref{fig:HFStiming}).

\begin{figure*}[t]
\centering
\includegraphics[width=7in]{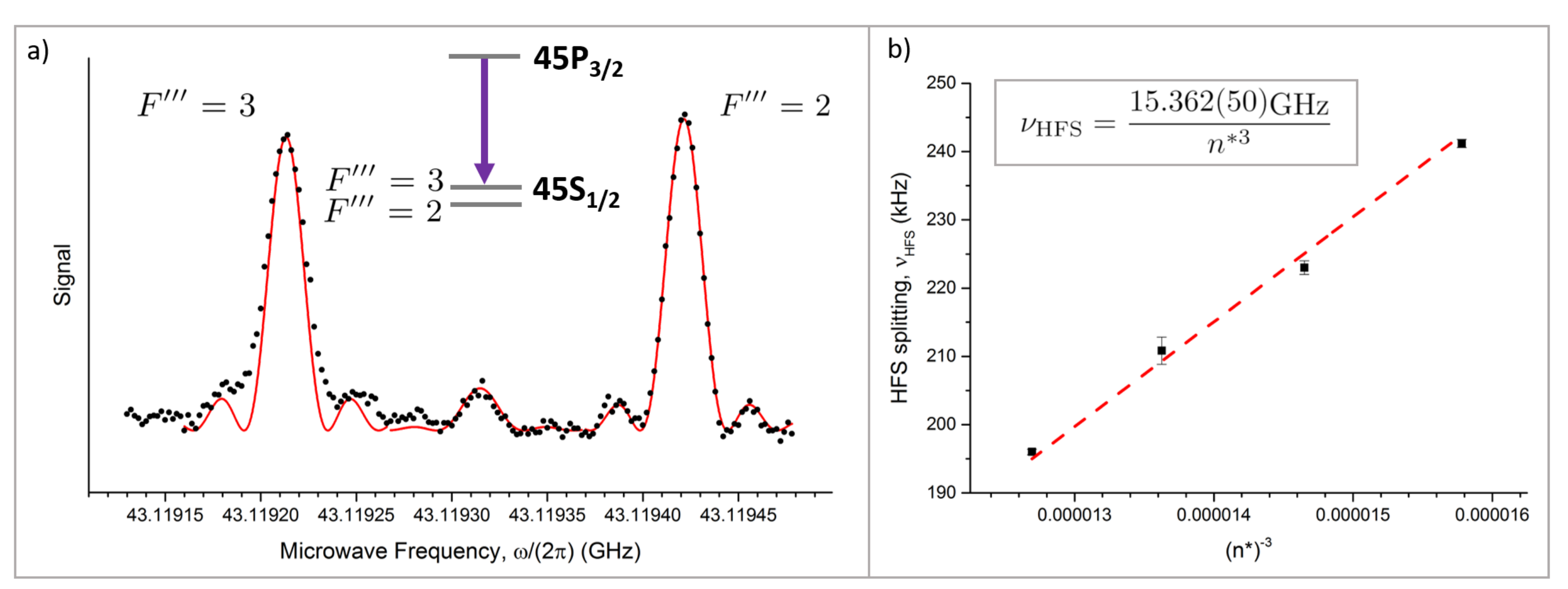}
\caption{a) Scan of the HFS of the $45P_{3/2} \rightarrow 45S_{1/2}$ transition (average over 6 scans with 300 experimental cycles per frequency point per scan). The frequency step size is 2~kHz. The red lines show fit results based on Eq.~\ref{eq:SaturatedSincSquared}, which yield the hyperfine line centers. The two largest peaks are the $F'''= 3$ and $F'''= 2$ HFS peaks of the $45S_{1/2}$ state. The smaller peak in the middle is thought to originate from transitions into hyperfine-mixed pair Rydberg states (see Discussion). b) HFS splittings as a function of $n^{*-3}$, where the effective principal quantum number $n^{*}= n-\delta_s(n)$. The dashed red line shows a linear fit to the data obtained with Origin software. The fitted slope and its uncertainty lead to the $A_{\mathrm{HFS}}$ value shown in the boxed equation.}
\label{fig:HFSsplittings}
\vspace{-0.1 in}
\end{figure*}

The population in the final Rydberg state is measured by applying an electric-field ionization ramp which ionizes the Rydberg atoms and accelerates the resulting ions towards a micro-channel plate (MCP). The $nP_{3/2}$ state ionizes first (at lower electric fields), closely followed by the $nS_{1/2}$ state~\cite{GallagherBook}. To determine the target-state population, we use two photon-counter gates, one only for $nS_{1/2}$ and the other for all possible states. We record the percent of population in the $nS_{1/2}$ state as a function of the scanned microwave frequency.

To coherently drive the microwave transitions of interest and to minimize their static-field level shifts, it is critical to carefully zero the magnetic and electric fields in the excitation area. The stray magnetic fields are zeroed down to $\leq$ 2.5~mG by changing the currents in three pairs of orthogonal compensation coils located around the vacuum chamber and observing the resulting Zeeman shifts of microwave transitions between Rydberg states.  As seen in Fig.~\ref{fig:HFScalculations} below, the $nP_{3/2} \rightarrow nS_{1/2}$ transitions have comparable HFS and Zeeman shifts. This complicates the spectra in applied magnetic fields, which renders the $nP_{3/2} \rightarrow nS_{1/2}$ transitions less useful to minimize stray magnetic fields. Therefore, for the magnetic field zeroing, we use the $54P_{3/2} \rightarrow 55D_{5/2}$ microwave transition, which is virtually independent of HFS effects and yields a clean Zeeman structure that can be easily analyzed to minimize stray magnetic fields.

We zero stray electric fields down to 0.2~mV/cm through the use of three pairs of electrodes that generate homogeneous orthogonal electric fields. The electric field is zeroed by first performing Stark spectroscopy~\cite{CisternasStarkSpec} on $nF$ Rydberg states, which have large electric polarizabilities. The peak locations of the $nF$ Rydberg lines are measured as a function of voltages applied to the in-vacuum electrodes, one direction at a time. The electric field is minimal when the $nF$ Rydberg lines are most blue-shifted. The accuracy of this method is limited by the linewidth of the optical-excitation Rydberg lines (typically about 4~MHz). A finer zeroing of the electric field is subsequently performed by analyzing the Stark shifts of microwave transitions between Rydberg levels. Since the linewidths of microwave transitions are narrower (for our case, typically tens of kHz), we are able to attain better Stark-shift resolution and hence cancel stray electric fields more precisely than with optical Stark spectroscopy alone.

\section{Results}

We are able to obtain Fourier-limited peaks, which, for our pulse width of $\tau =$ 40~$\mu$s, have linewidths of 22~kHz (see Fig.~\ref{fig:HFSsplittings}). In the scans, several Fourier sidebands are also resolved, with the expected periodicity of $\omega \tau = n\pi$, where $\omega$ is the microwave angular frequency. The presence of Fourier side peaks is an indicator of coherent transitions and the absence of inhomogeneous broadening effects. Furthermore, our observation of coherent transitions indicates that the HFS splitting of the $nP_{3/2}$ level only plays an insignificant role (see Discussion).

In order to obtain precise measurements of the HFS splittings, it is critical to fit the observed data to the appropriate lineshapes, and to extract the center frequencies of each peak. The lineshape expected for a square pulse that saturates the excitation follows~\cite{Meystre2007, BermanBook}
\begin{equation}
\begin{split}
\label{eq:SaturatedSincSquared}
P_e &= A\left [ \frac{\Omega^2}{\Omega^2+(\omega-\omega_c)^2} \right ] \\
&\quad  \times \mathrm{sin}^2\left [ \frac{1}{2} \tau \sqrt{\Omega^2+(\omega-\omega_c)^2} \right ] ,
\end{split}
\end{equation}
\noindent where $P_e$ is the excited-state population, $A$ is an amplitude parameter, $\Omega$ is the on-resonance angular Rabi frequency, $\omega_c$ is the main peak's center angular frequency and $\tau$ is the interaction time, which for these experiments is fixed at 40~$\mu s$. We use Eq.~\ref{eq:SaturatedSincSquared} as a fit function to determine the line center for each peak in our data sets.

A sample set of results for $n=45$ is shown in Fig.~\ref{fig:HFSsplittings}a. The data points shown are averages over 6 scans, each with 300 experimental cycles per frequency point per scan. The red curves show fits based on Eq.~\ref{eq:SaturatedSincSquared}. For the other $n$-states (shown in Fig.~\ref{fig:HFSsplittings}b), we average over 3-4 scans, each with 300 experimental cycles as well.

In Table~\ref{tab:HFSsplittings} we list the extracted HFS splittings for $n=$ 43, 44, 45 and 46 with their respective statistical uncertainties. To obtain the HFS splittings, we fit the two main peaks in each of the 3-6 scans taken for a given principal quantum number $n$ to determine the line centers. For each $n$, we then record the difference between the $F'''= 2$ and $F'''= 3$ line centers and take the average of these differences. To arrive at the uncertainty of these splittings, we first determine the uncertainty of the line centers of each peak ($F'''= 2$ and $F'''= 3$) by taking the standard error of the mean of the fitted line centers. We then proceed to propagate the error by adding the uncertainties from each peak in quadrature, which yields the uncertainty in the splitting for each of the principal quantum numbers, $n$. From the obtained HFS splittings, the individual $A_{\mathrm{HFS},n}$, shown in Table~\ref{tab:HFSsplittings}, are extracted using Eq.~\ref{eq:HFSsplitting1}. Their uncertainties are determined by propagating the uncertainty in the splitting using Eq.~\ref{eq:HFSsplitting1}.

\begin{table}[h]
\centering
\caption{\label{tab:HFSsplittings} Measured HFS splittings for different $nP_{3/2} \rightarrow nS_{1/2}$ transitions of $^{85}$Rb and their respective hyperfine coupling constants, $A_{\mathrm{HFS},n}$. Only statistical uncertainties are displayed.}
\begin{tabular}{ccc}
\\
$n$ & HFS splitting (kHz) & $A_{\mathrm{HFS},n}$ (GHz) \\
\hline
\hline
43 & 241.2(6) &	15.284(39) \\
44 & 223(1) &	15.222(77) \\
45 & 211(2) &   15.47(12) \\
46 & 196.0(4) &   15.440(30) \\
\end{tabular}
\end{table}

\begin{table}[b]
\centering
\caption{\label{tab:HFSbudget} Uncertainty budget for the $A_{\mathrm{HFS}}$ measurement of $^{85}$Rb. Values shown for the systematic uncertainties were obtained from the $45P_{3/2} \rightarrow 45S_{1/2}$ transition. Statistical uncertainty includes data for all $n$.}
\begin{tabular}{lc}
\\
Source & Uncertainty in $A_{\mathrm{HFS}}$ (MHz) \\
\hline
\hline
Magnetic field & 52 \\
Rydberg interactions & 46 \\
Electric field & 33 \\
Statistical  &    23 \\
\end{tabular}
\end{table}

The final $A_{\mathrm{HFS}}$ displayed in Eq.~\ref{eq:HFSsplitting} is given by the weighted mean of all $A_{\mathrm{HFS}, n}$ in Table~\ref{tab:HFSsplittings},
\[ A_{\mathrm{HFS}} = \sum_{n} \frac{A_{\mathrm{HFS},n}}{\sigma_n^2} / \sum_{n} \frac{1}{\sigma_n^2} \quad, \]
where $\sigma_n$ is the uncertainty in $A_{\mathrm{HFS},n}$.
The statistical uncertainty of the weighted average ($\sqrt{1/\frac{1}{\sum_n \sigma_n^2}}$) is 23~MHz. In Eq.~\ref{eq:HFSsplitting}, the uncertainty in $\delta_s$ is negligible at the level of precision of the present measurements.

\begin{figure}[t]
\centering
\includegraphics[width=3in]{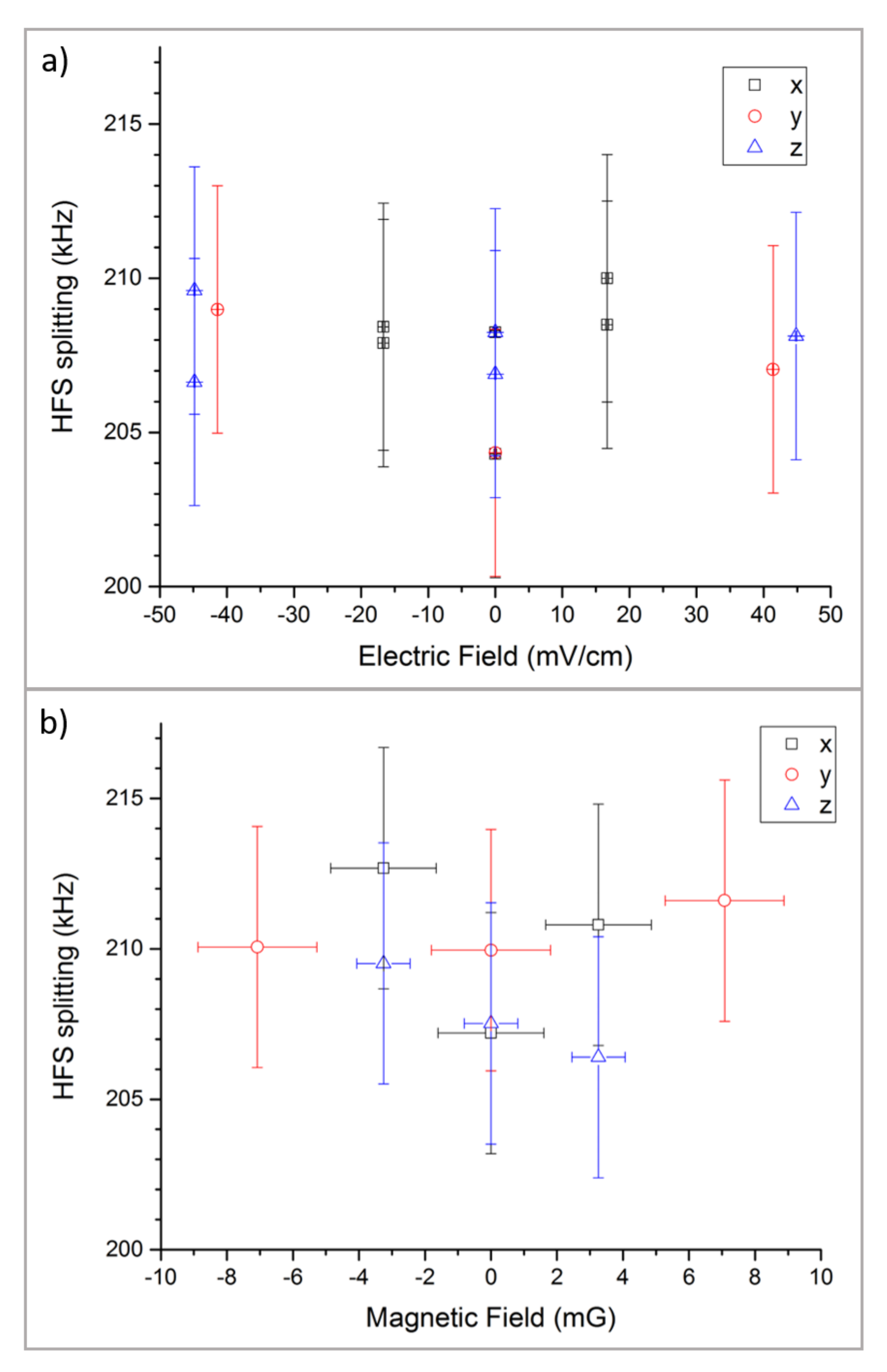}
\caption{(color online). Measured HFS splitting as the electric (a) and magnetic (b) fields are changed one direction at a time, with all other field components compensated to zero (within our uncertainties). All data were obtained for the $45P_{3/2} \rightarrow 45S_{1/2}$ microwave transition. Vertical error bars result from adding in quadrature the 1$\sigma$ uncertainties in the individual line centers. Horizontal error bars reflect
voltage and current uncertainties for (a) and (b), respectively.}
\label{fig:HFSsystematicFields}
\vspace{-0.1 in}
\end{figure}

Besides looking at statistical uncertainty, we also explore possible sources of systematic uncertainty. To study the dependence of the splitting on electric fields, we apply well-known electric fields in one direction at a time and record the HFS splitting at each electric-field step (see Fig.~\ref{fig:HFSsystematicFields}a). We find that there is no significant change in the splitting due to electric fields, meaning that the $F'''= 2$ and $F'''= 3$ peaks are shifted equally by stray electric fields within our statistical resolution. This finding agrees well with the behavior observed in~\cite{SpreeuwHFS}. We use the standard error of the mean of all data in Fig.~\ref{fig:HFSsystematicFields}a and Eq.~1 to determine an upper bound for the systematic uncertainty in $A_{\mathrm{HFS}}$ that may arise from stray electric field effects; this upper bound is listed in Table~\ref{tab:HFSbudget}.

\begin{figure}[t]
\includegraphics[width=3in]{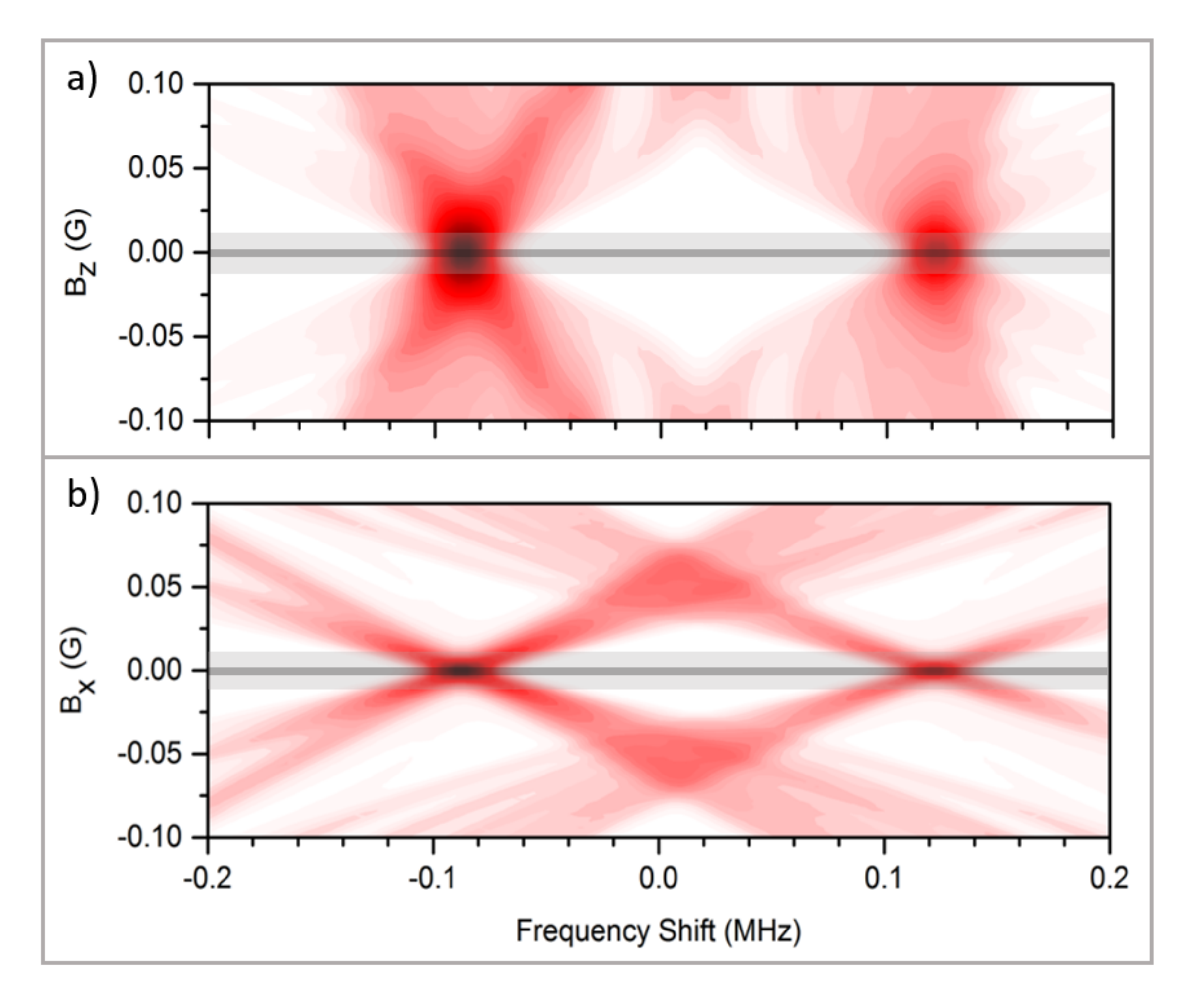}
\caption{(color online). Calculations of the $45P_{3/2} \rightarrow 45S_{1/2}$ $F'''= 3$ (left peak) and $F'''= 2$ (right peak) hyperfine peaks as a function of a parallel (a) and a transverse (b) magnetic field. Highest final-state population is shown in dark red, while white is zero. The dark gray horizontal bar shows the uncertainty in our magnetic-field zeroing (2.5~mG), while the light gray horizontal bar shows the range of magnetic fields applied to obtain the data in Fig.~\ref{fig:HFSsystematicFields}b.}
\label{fig:HFScalculations}
\vspace{-0.1 in}
\end{figure}

To explore the effects of magnetic fields, we record the HFS splitting as a function of applied magnetic field, one direction at a time (see Fig.~\ref{fig:HFSsystematicFields}b). We find that within $\pm$ 10~mG there is no significant change in the measured splitting. This is in good agreement with calculations shown in Fig.~\ref{fig:HFScalculations}, where we diagonalize the Hamiltonian $\hat{H}= \hat{H}_{\mathrm{HFS}}+ \hat{H}_{B}$. Here, $\hat{H}_{\mathrm{HFS}}$ is the HFS interaction and $\hat{H}_{B}$ the Zeeman interaction, for a magnetic field $B$ that is parallel or transverse with respect to the incident $\pi$-polarized microwaves. We use the $A_{\mathrm{HFS}}$ value for $nS$-states we obtain from our experiments (see Eq.~\ref{eq:HFSsplitting}), and we assume that the $nP$-states have zero HFS splitting. The line strength information required to prepare Fig.~\ref{fig:HFScalculations} is obtained by calculating electric-dipole matrix elements between the eigenstates of $\hat{H}$ in the applied magnetic field, for $\pi$-polarized microwaves. A detailed analysis of the data in Fig.~\ref{fig:HFScalculations} shows that within $\pm$ 10~mG the splitting changes by at most 2~kHz, while the lines broaden to about 50~kHz at $\pm$ 10~mG. The rapid broadening of the calculated HFS lines in Fig.~\ref{fig:HFScalculations} as a function of magnetic field shows that the linewidth of the HFS transitions is a sensitive measure for the stray magnetic field. The absence of significant inhomogeneous broadening in the experiment represents a valuable secondary confirmation that the magnetic field has indeed an upper limit of a few mG. We use the standard error of the mean of the data in Fig.~\ref{fig:HFSsystematicFields}b to find an upper bound of the stray-magnetic-field-induced systematic uncertainty in $A_{\mathrm{HFS}}$, listed in Table~\ref{tab:HFSbudget}.


\begin{figure}[b]
\centering
\includegraphics[width=3.4in]{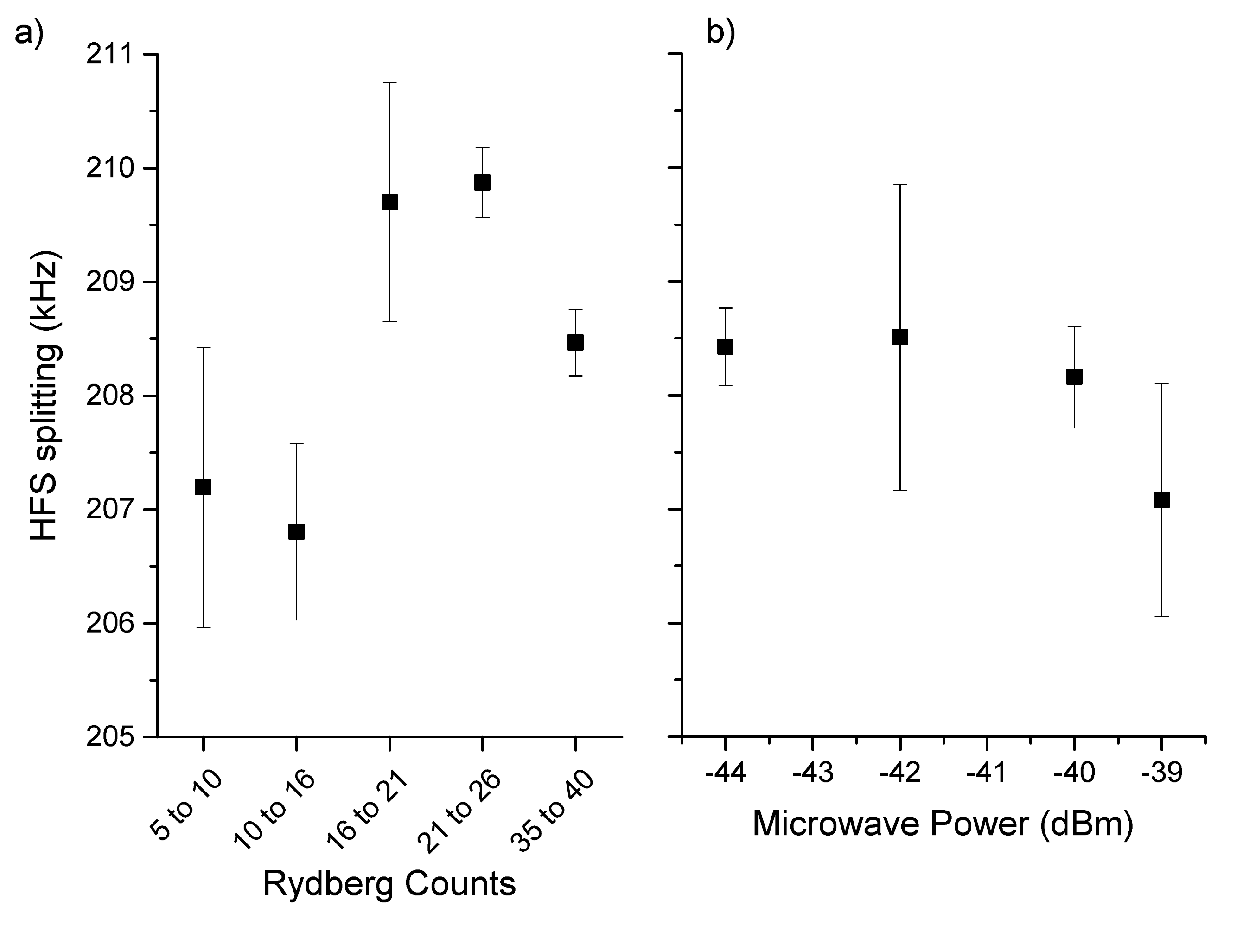}
\caption{Measured hyperfine structure splitting for different number of Rydberg counts (a) and different microwave generator output powers (b). All data were obtained for the $45P_{3/2} \rightarrow 45S_{1/2}$ microwave transition. Each data point is the average of 2-4 scans and the vertical error bar is the standard error of the mean.}
\label{fig:HFSsystematics}
\vspace{-0.1 in}
\end{figure}

Another possible concern in our measurements are dipolar and van der Waals Rydberg-Rydberg interactions, which scale as $R^{-3}$~\cite{GallagherDipole} and $R^{-6}$~\cite{Reinhard2007}, respectively ($R$ is the inter-atomic distance). To investigate this, we perform measurements for $n=45$ as a function of the number of detected Rydberg atoms (see Fig.~\ref{fig:HFSsystematics}a). We find that the hyperfine splitting varies over a full range of about $\pm$ 3~kHz, equivalent to $\pm$ 0.7$\%$. We believe that the variation seen at Rydberg counts $\gtrsim 16$ is due to dipole-dipole interactions, which occur due to the presence of both $S$ and $P$ Rydberg atoms in our spectroscopic sample. We estimate the dipolar interactions to produce a shift on the order of a few kHz. To arrive at this value, we estimate $R$ from the number of Rydberg counts, the MCP ion-detection efficiency ($30\%$) and the interaction volume ($\sim 1$~mm$^3$). In Fig.~\ref{fig:HFSsystematics}a, the Rydberg-atom density ranges up to $n_V \sim 3.5 \times 10^4$ cm$^{-3}$, corresponding to $R \gtrsim$ 100~$\mu$m. Noting that the dipole-dipole interaction scales as $(n^2ea_0)^2/(4\pi h \epsilon_0 R^3)$, we arrive at a dipole-dipole shift of $\sim$ 4~kHz. This value matches well with the variation observed in Fig.~\ref{fig:HFSsystematics}a. The measurements in Table~\ref{tab:HFSsplittings} were all done in the low Rydberg-atom density regime (16 Rydberg counts and below). For an upper bound of the systematic uncertainty in $A_{\mathrm{HFS}}$ that may arise from dipolar Rydberg-Rydberg interactions, listed in Table~\ref{tab:HFSbudget}, we use the standard error of the mean of all data in Fig.~\ref{fig:HFSsystematics}a.

For completeness, we also estimate the van der Waals interaction strength, which is on the order of $6 \, C_6 n_V^{2}$, where $C_6$ is the van der Waals coefficient and $n_V$ the Rydberg-atom density. In~\cite{Pound2015} it was found that $C_6 = 5.4 \times 10^{-58}$~Jm$^6$ for 70$S_{1/2}$, in agreement with~\cite{Reinhard2007}. Noting that $C_6$ scales approximately as $n^{*11}$, with effective quantum number $n^*$, for the $45S_{1/2}$ state it is $C_6 = 3 \times 10^{-60}$~Jm$^6$, and for the above maximum density the van der Waals shift is $\sim$1~mHz, which is negligible. For 45$P_{3/2}$ we have calculated Rydberg pair potentials~\cite{Sassmannshausen2015, Sassmannshausen2016, Cote2002} and found $C_6$ coefficients ranging between $-2.5 \times 10^{-59}$~Jm$^6$ and $+7 \times 10^{-61}$~Jm$^6$ (depending on angular-momentum projections on the internuclear axis).  The corresponding van der Waals shifts have magnitudes less than 10~mHz and are also negligible.

We also have considered the AC-Stark shifts caused by the microwave used to drive the $nP_{3/2} \rightarrow nS_{1/2}$ transitions. To this end, we measure the hyperfine structure splitting for different microwave powers and observe no significant difference over the range of powers tested (see Fig.~\ref{fig:HFSsystematics}b). Based on lineshape fits analogous to the ones shown in Fig.~\ref{fig:HFSsplittings}, we estimate that for a microwave-source output power of -39~dBm the Rabi frequency of the atoms is 8-9~kHz. Using this value, calculated matrix elements of about 1000~$ea_0$ and a calculated AC polarizability of the transition of $\lesssim$ 1~MHz/(V/m)$^2$ over the principal-quantum-number range $n =$~43-46, we estimate sub-Hertz AC transition shifts. Since the estimated AC shifts are more than three orders of magnitude below the dominant uncertainties, they can be safely ignored.



Taking all uncertainties listed in Table~\ref{tab:HFSbudget} into account, the resulting expression for the $nS_{1/2}$ hyperfine splitting is
\begin{equation}
\label{eq:HFSsplitting}
\nu_{\mathrm{HFS}}= 15.372(80)~\mathrm{GHz}~  \frac{1}{[n-\delta_s(n)]^3},
\end{equation}
\noindent where $\delta_s(n)$ is the quantum defect of $nS$-states obtained from~\cite{GallagherNsNp}. This leads us to our final result, $A_{\mathrm{HFS}}=$15.372(80)~GHz.  

As a consistency check, we compare our result with
the linear-fit result shown in Fig.~\ref{fig:HFSsplittings}b, which is $A_{\mathrm{HFS}}=$15.362(50)~GHz, and note good agreement. Our result also is consistent with a previous measurement~\cite{GallagherNsNp}, which yielded a value of 14.6(14)~GHz.

\section{Discussion}

In our analysis, the hyperfine structure splitting of the $nP_{3/2}$ state is taken to be negligible. As a justification, we calculate the hyperfine splittings of $45P_{3/2}$ using the hyperfine $A$ and $B$ constants given in~\cite{Arimondo1977} (which are for $n\leq 10$ but scale close to $n^{*-3}$). Since we populate $nP_{3/2}, F''=$ 2, 3 and 4, there are a total of 5 possible hyperfine transitions, two to $nS_{1/2}, F'''=$ 2 separated by about 10~kHz, and three to $nS_{1/2}, F'''=$ 3 separated also by about 10~kHz from one another. Variations in the line strengths of these hyperfine transitions, caused by laser-polarization and ground-state magnetization drifts, may affect the net line shapes, and hence the measured $nS_{1/2}$ hyperfine splittings. As shown in Fig.~\ref{fig:HFSsplittings}, the $45S_{1/2}, F'''= 2$ peak has a larger amplitude than the $45S_{1/2}, F'''= 3$ peak. The same behavior is observed for the $n =$ 43, 44 and 46 cases. This leads us to believe that the $nP_{3/2}, F''= 4$ is not strongly populated. In that case, there are two unresolved sub-peaks under each $nS_{1/2}, F'''$ hyperfine component. The splitting of the unresolved sub-peaks is $\approx 10$~kHz, which is less than the $\sim 40$-kHz splitting of the well-resolved Fourier sidebands in Fig.~\ref{fig:HFSsplittings}. We therefore conclude that the hyperfine structure of the $nP_{3/2}$ state does not have an effect on our measurements at the present level of precision.


Figure~\ref{fig:HFSsplittings} also shows a central feature mid-way between the hyperfine peaks. This central feature also consistently appears in the scans for all other $n$, it is Fourier-limited, and has Fourier sidebands. We
believe that this feature is due to the fact that the atoms are at close-enough separations to allow
for simultaneous microwave transitions of atom pairs in $45P_{3/2}$ into symmetrized hyperfine-mixed Rydberg pair states of the type $\vert 45S_{1/2}, F'''=2 \rangle_A \otimes
\vert 45S_{1/2}, F'''=3 \rangle_B + $ $\vert 45S_{1/2}, F'''=3 \rangle_A \otimes
\vert 45S_{1/2}, F'''=2 \rangle_B$, with $A$ and $B$ referring to two Rydberg atoms at center-of-mass positions ${\bf{R}}_A$ and ${\bf{R}}_B$. This is a two-photon microwave transition in the product space of atoms $A$ and $B$.
Signals due to transitions into such hyperfine-mixed $(45S_{1/2})_2$ pair states would appear exactly halfway between the $F'''= 3$ and $F'''= 2$ peaks, as is the case for the features we observe. Further analysis of this observation is of considerable interest.

In conclusion, we have measured the hyperfine structure splittings of $nS_{1/2}$ states of $^{85}$Rb and extracted a hyperfine coupling constant that is an order of magnitude more precise than the best previously attained measurement~\cite{GallagherNsNp}, and is consistent with that measurement. All known systematic effects were included in our data analysis. The data also provide some evidence for a new type of hyperfine-mixed Rydberg pair states.

\section{Acknowledgments}

A.R. acknowledges support from the National Science Foundation Graduate Research Fellowship under Grant No. DGE 1256260 and the University of Michigan Rackham Pre-doctoral Fellowship. This work was supported by NSF Grant No. PHY1806809, and NASA Grant No. NNH13ZTT002N NRA.


\bibliographystyle{apsrev4-1}
\bibliography{HFS_Bibliography}

\providecommand{\noopsort}[1]{}\providecommand{\singleletter}[1]{#1}%
\begin{thebibliography}{26}%
\makeatletter
\providecommand \@ifxundefined [1]{%
 \@ifx{#1\undefined}
}%
\providecommand \@ifnum [1]{%
 \ifnum #1\expandafter \@firstoftwo
 \else \expandafter \@secondoftwo
 \fi
}%
\providecommand \@ifx [1]{%
 \ifx #1\expandafter \@firstoftwo
 \else \expandafter \@secondoftwo
 \fi
}%
\providecommand \natexlab [1]{#1}%
\providecommand \enquote  [1]{``#1''}%
\providecommand \bibnamefont  [1]{#1}%
\providecommand \bibfnamefont [1]{#1}%
\providecommand \citenamefont [1]{#1}%
\providecommand \href@noop [0]{\@secondoftwo}%
\providecommand \href [0]{\begingroup \@sanitize@url \@href}%
\providecommand \@href[1]{\@@startlink{#1}\@@href}%
\providecommand \@@href[1]{\endgroup#1\@@endlink}%
\providecommand \@sanitize@url [0]{\catcode `\\12\catcode `\$12\catcode
  `\&12\catcode `\#12\catcode `\^12\catcode `\_12\catcode `\%12\relax}%
\providecommand \@@startlink[1]{}%
\providecommand \@@endlink[0]{}%
\providecommand \url  [0]{\begingroup\@sanitize@url \@url }%
\providecommand \@url [1]{\endgroup\@href {#1}{\urlprefix }}%
\providecommand \urlprefix  [0]{URL }%
\providecommand \Eprint [0]{\href }%
\providecommand \doibase [0]{http://dx.doi.org/}%
\providecommand \selectlanguage [0]{\@gobble}%
\providecommand \bibinfo  [0]{\@secondoftwo}%
\providecommand \bibfield  [0]{\@secondoftwo}%
\providecommand \translation [1]{[#1]}%
\providecommand \BibitemOpen [0]{}%
\providecommand \bibitemStop [0]{}%
\providecommand \bibitemNoStop [0]{.\EOS\space}%
\providecommand \EOS [0]{\spacefactor3000\relax}%
\providecommand \BibitemShut  [1]{\csname bibitem#1\endcsname}%
\let\auto@bib@innerbib\@empty
\bibitem [{\citenamefont {Steck}(2008)}]{Steck85Rb}%
  \BibitemOpen
  \bibfield  {author} {\bibinfo {author} {\bibfnamefont {D.}~\bibnamefont
  {Steck}},\ }\href@noop {} {\enquote {\bibinfo {title} {Rubidium 85 d line
  data},}\ } (\bibinfo {year} {2008})\BibitemShut {NoStop}%
\bibitem [{\citenamefont {Arimondo}\ \emph {et~al.}(1977)\citenamefont
  {Arimondo}, \citenamefont {Inguscio},\ and\ \citenamefont
  {Violino}}]{Arimondo1977}%
  \BibitemOpen
  \bibfield  {author} {\bibinfo {author} {\bibfnamefont {E.}~\bibnamefont
  {Arimondo}}, \bibinfo {author} {\bibfnamefont {M.}~\bibnamefont {Inguscio}},
  \ and\ \bibinfo {author} {\bibfnamefont {P.}~\bibnamefont {Violino}},\ }\href
  {\doibase 10.1103/RevModPhys.49.31} {\bibfield  {journal} {\bibinfo
  {journal} {Rev. Mod. Phys.}\ }\textbf {\bibinfo {volume} {49}},\ \bibinfo
  {pages} {31} (\bibinfo {year} {1977})}\BibitemShut {NoStop}%
\bibitem [{\citenamefont {Ladd}\ \emph {et~al.}(2010)\citenamefont {Ladd},
  \citenamefont {Jelezko}, \citenamefont {Laflamme}, \citenamefont {Nakamura},
  \citenamefont {Monroe},\ and\ \citenamefont {O’Brien}}]{Ladd2010}%
  \BibitemOpen
  \bibfield  {author} {\bibinfo {author} {\bibfnamefont {T.~D.}\ \bibnamefont
  {Ladd}}, \bibinfo {author} {\bibfnamefont {F.}~\bibnamefont {Jelezko}},
  \bibinfo {author} {\bibfnamefont {R.}~\bibnamefont {Laflamme}}, \bibinfo
  {author} {\bibfnamefont {Y.}~\bibnamefont {Nakamura}}, \bibinfo {author}
  {\bibfnamefont {C.}~\bibnamefont {Monroe}}, \ and\ \bibinfo {author}
  {\bibfnamefont {J.~L.}\ \bibnamefont {O’Brien}},\ }\href@noop {} {\bibfield
   {journal} {\bibinfo  {journal} {Nature}\ }\textbf {\bibinfo {volume}
  {464}},\ \bibinfo {pages} {45} (\bibinfo {year} {2010})}\BibitemShut
  {NoStop}%
\bibitem [{\citenamefont {Dumke}\ \emph {et~al.}(2002)\citenamefont {Dumke},
  \citenamefont {Volk}, \citenamefont {M{\"u}ther}, \citenamefont {Buchkremer},
  \citenamefont {Birkl},\ and\ \citenamefont {Ertmer}}]{HFSquantumCompRb2002}%
  \BibitemOpen
  \bibfield  {author} {\bibinfo {author} {\bibfnamefont {R.}~\bibnamefont
  {Dumke}}, \bibinfo {author} {\bibfnamefont {M.}~\bibnamefont {Volk}},
  \bibinfo {author} {\bibfnamefont {T.}~\bibnamefont {M{\"u}ther}}, \bibinfo
  {author} {\bibfnamefont {F.~B.~J.}\ \bibnamefont {Buchkremer}}, \bibinfo
  {author} {\bibfnamefont {G.}~\bibnamefont {Birkl}}, \ and\ \bibinfo {author}
  {\bibfnamefont {W.}~\bibnamefont {Ertmer}},\ }\href@noop {} {\bibfield
  {journal} {\bibinfo  {journal} {Phys. Rev. Lett.}\ }\textbf {\bibinfo
  {volume} {89}},\ \bibinfo {pages} {097903} (\bibinfo {year}
  {2002})}\BibitemShut {NoStop}%
\bibitem [{\citenamefont {Harty}\ \emph {et~al.}(2014)\citenamefont {Harty},
  \citenamefont {Allcock}, \citenamefont {Ballance}, \citenamefont {Guidoni},
  \citenamefont {Janacek}, \citenamefont {Linke}, \citenamefont {Stacey},\ and\
  \citenamefont {Lucas}}]{HFSquantumCompIons2014}%
  \BibitemOpen
  \bibfield  {author} {\bibinfo {author} {\bibfnamefont {T.}~\bibnamefont
  {Harty}}, \bibinfo {author} {\bibfnamefont {D.}~\bibnamefont {Allcock}},
  \bibinfo {author} {\bibfnamefont {C.~J.}\ \bibnamefont {Ballance}}, \bibinfo
  {author} {\bibfnamefont {L.}~\bibnamefont {Guidoni}}, \bibinfo {author}
  {\bibfnamefont {H.}~\bibnamefont {Janacek}}, \bibinfo {author} {\bibfnamefont
  {N.}~\bibnamefont {Linke}}, \bibinfo {author} {\bibfnamefont
  {D.}~\bibnamefont {Stacey}}, \ and\ \bibinfo {author} {\bibfnamefont
  {D.}~\bibnamefont {Lucas}},\ }\href@noop {} {\bibfield  {journal} {\bibinfo
  {journal} {Phys. Rev. Lett.}\ }\textbf {\bibinfo {volume} {113}},\ \bibinfo
  {pages} {220501} (\bibinfo {year} {2014})}\BibitemShut {NoStop}%
\bibitem [{\citenamefont {Lukin}\ \emph {et~al.}(2001)\citenamefont {Lukin},
  \citenamefont {Fleischhauer}, \citenamefont {Cote}, \citenamefont {Duan},
  \citenamefont {Jaksch}, \citenamefont {Cirac},\ and\ \citenamefont
  {Zoller}}]{Lukin2001}%
  \BibitemOpen
  \bibfield  {author} {\bibinfo {author} {\bibfnamefont {M.}~\bibnamefont
  {Lukin}}, \bibinfo {author} {\bibfnamefont {M.}~\bibnamefont {Fleischhauer}},
  \bibinfo {author} {\bibfnamefont {R.}~\bibnamefont {Cote}}, \bibinfo {author}
  {\bibfnamefont {L.}~\bibnamefont {Duan}}, \bibinfo {author} {\bibfnamefont
  {D.}~\bibnamefont {Jaksch}}, \bibinfo {author} {\bibfnamefont
  {J.}~\bibnamefont {Cirac}}, \ and\ \bibinfo {author} {\bibfnamefont
  {P.}~\bibnamefont {Zoller}},\ }\href@noop {} {\bibfield  {journal} {\bibinfo
  {journal} {Phys. Rev. Lett.}\ }\textbf {\bibinfo {volume} {87}},\ \bibinfo
  {pages} {037901} (\bibinfo {year} {2001})}\BibitemShut {NoStop}%
\bibitem [{\citenamefont {Saffman}\ \emph {et~al.}(2010)\citenamefont
  {Saffman}, \citenamefont {Walker},\ and\ \citenamefont
  {M{\o}lmer}}]{SaffmanReview2010}%
  \BibitemOpen
  \bibfield  {author} {\bibinfo {author} {\bibfnamefont {M.}~\bibnamefont
  {Saffman}}, \bibinfo {author} {\bibfnamefont {T.~G.}\ \bibnamefont {Walker}},
  \ and\ \bibinfo {author} {\bibfnamefont {K.}~\bibnamefont {M{\o}lmer}},\
  }\href@noop {} {\bibfield  {journal} {\bibinfo  {journal} {Rev. Mod. Phys.}\
  }\textbf {\bibinfo {volume} {82}},\ \bibinfo {pages} {2313} (\bibinfo {year}
  {2010})}\BibitemShut {NoStop}%
\bibitem [{\citenamefont {Saffman}(2016)}]{SaffmanReview2016}%
  \BibitemOpen
  \bibfield  {author} {\bibinfo {author} {\bibfnamefont {M.}~\bibnamefont
  {Saffman}},\ }\href@noop {} {\bibfield  {journal} {\bibinfo  {journal} {J.
  Phys. B: At. Mol. Opt. Phys.}\ }\textbf {\bibinfo {volume} {49}},\ \bibinfo
  {pages} {202001} (\bibinfo {year} {2016})}\BibitemShut {NoStop}%
\bibitem [{\citenamefont {Safronova}\ \emph {et~al.}(2018)\citenamefont
  {Safronova}, \citenamefont {Budker}, \citenamefont {DeMille}, \citenamefont
  {Kimball}, \citenamefont {Derevianko},\ and\ \citenamefont
  {Clark}}]{SafronovaReview2018}%
  \BibitemOpen
  \bibfield  {author} {\bibinfo {author} {\bibfnamefont {M.}~\bibnamefont
  {Safronova}}, \bibinfo {author} {\bibfnamefont {D.}~\bibnamefont {Budker}},
  \bibinfo {author} {\bibfnamefont {D.}~\bibnamefont {DeMille}}, \bibinfo
  {author} {\bibfnamefont {D.~F.~J.}\ \bibnamefont {Kimball}}, \bibinfo
  {author} {\bibfnamefont {A.}~\bibnamefont {Derevianko}}, \ and\ \bibinfo
  {author} {\bibfnamefont {C.~W.}\ \bibnamefont {Clark}},\ }\href@noop {}
  {\bibfield  {journal} {\bibinfo  {journal} {Rev. Mod. Phys.}\ }\textbf
  {\bibinfo {volume} {90}},\ \bibinfo {pages} {025008} (\bibinfo {year}
  {2018})}\BibitemShut {NoStop}%
\bibitem [{\citenamefont {Anderson}\ \emph {et~al.}(2014)\citenamefont
  {Anderson}, \citenamefont {Miller},\ and\ \citenamefont
  {Raithel}}]{Anderson2014}%
  \BibitemOpen
  \bibfield  {author} {\bibinfo {author} {\bibfnamefont {D.~A.}\ \bibnamefont
  {Anderson}}, \bibinfo {author} {\bibfnamefont {S.~A.}\ \bibnamefont
  {Miller}}, \ and\ \bibinfo {author} {\bibfnamefont {G.}~\bibnamefont
  {Raithel}},\ }\href@noop {} {\bibfield  {journal} {\bibinfo  {journal} {Phys.
  Rev. A}\ }\textbf {\bibinfo {volume} {90}},\ \bibinfo {pages} {062518}
  (\bibinfo {year} {2014})}\BibitemShut {NoStop}%
\bibitem [{\citenamefont {Sa{\ss}mannshausen}\ \emph
  {et~al.}(2015)\citenamefont {Sa{\ss}mannshausen}, \citenamefont {Merkt},\
  and\ \citenamefont {Deiglmayr}}]{Sassmannshausen2015}%
  \BibitemOpen
  \bibfield  {author} {\bibinfo {author} {\bibfnamefont {H.}~\bibnamefont
  {Sa{\ss}mannshausen}}, \bibinfo {author} {\bibfnamefont {F.}~\bibnamefont
  {Merkt}}, \ and\ \bibinfo {author} {\bibfnamefont {J.}~\bibnamefont
  {Deiglmayr}},\ }\href@noop {} {\bibfield  {journal} {\bibinfo  {journal}
  {Phys. Rev. Lett.}\ }\textbf {\bibinfo {volume} {114}},\ \bibinfo {pages}
  {133201} (\bibinfo {year} {2015})}\BibitemShut {NoStop}%
\bibitem [{\citenamefont {B{\"o}ttcher}\ \emph {et~al.}(2016)\citenamefont
  {B{\"o}ttcher}, \citenamefont {Gaj}, \citenamefont {Westphal}, \citenamefont
  {Schlagm{\"u}ller}, \citenamefont {Kleinbach}, \citenamefont {L{\"o}w},
  \citenamefont {Liebisch}, \citenamefont {Pfau},\ and\ \citenamefont
  {Hofferberth}}]{Bottcher2016}%
  \BibitemOpen
  \bibfield  {author} {\bibinfo {author} {\bibfnamefont {F.}~\bibnamefont
  {B{\"o}ttcher}}, \bibinfo {author} {\bibfnamefont {A.}~\bibnamefont {Gaj}},
  \bibinfo {author} {\bibfnamefont {K.~M.}\ \bibnamefont {Westphal}}, \bibinfo
  {author} {\bibfnamefont {M.}~\bibnamefont {Schlagm{\"u}ller}}, \bibinfo
  {author} {\bibfnamefont {K.~S.}\ \bibnamefont {Kleinbach}}, \bibinfo {author}
  {\bibfnamefont {R.}~\bibnamefont {L{\"o}w}}, \bibinfo {author} {\bibfnamefont
  {T.~C.}\ \bibnamefont {Liebisch}}, \bibinfo {author} {\bibfnamefont
  {T.}~\bibnamefont {Pfau}}, \ and\ \bibinfo {author} {\bibfnamefont
  {S.}~\bibnamefont {Hofferberth}},\ }\href@noop {} {\bibfield  {journal}
  {\bibinfo  {journal} {Phys. Rev. A}\ }\textbf {\bibinfo {volume} {93}},\
  \bibinfo {pages} {032512} (\bibinfo {year} {2016})}\BibitemShut {NoStop}%
\bibitem [{\citenamefont {MacLennan}\ \emph {et~al.}(2019)\citenamefont
  {MacLennan}, \citenamefont {Chen},\ and\ \citenamefont
  {Raithel}}]{Maclennan2019}%
  \BibitemOpen
  \bibfield  {author} {\bibinfo {author} {\bibfnamefont {J.~L.}\ \bibnamefont
  {MacLennan}}, \bibinfo {author} {\bibfnamefont {Y.-J.}\ \bibnamefont {Chen}},
  \ and\ \bibinfo {author} {\bibfnamefont {G.}~\bibnamefont {Raithel}},\
  }\href@noop {} {\bibfield  {journal} {\bibinfo  {journal} {Phys. Rev. A}\
  }\textbf {\bibinfo {volume} {99}},\ \bibinfo {pages} {033407} (\bibinfo
  {year} {2019})}\BibitemShut {NoStop}%
\bibitem [{\citenamefont {Li}\ \emph {et~al.}(2003)\citenamefont {Li},
  \citenamefont {Mourachko}, \citenamefont {Noel},\ and\ \citenamefont
  {Gallagher}}]{GallagherNsNp}%
  \BibitemOpen
  \bibfield  {author} {\bibinfo {author} {\bibfnamefont {W.}~\bibnamefont
  {Li}}, \bibinfo {author} {\bibfnamefont {I.}~\bibnamefont {Mourachko}},
  \bibinfo {author} {\bibfnamefont {M.~W.}\ \bibnamefont {Noel}}, \ and\
  \bibinfo {author} {\bibfnamefont {T.~F.}\ \bibnamefont {Gallagher}},\ }\href
  {\doibase 10.1103/PhysRevA.67.052502} {\bibfield  {journal} {\bibinfo
  {journal} {Phys. Rev. A}\ }\textbf {\bibinfo {volume} {67}},\ \bibinfo
  {pages} {052502} (\bibinfo {year} {2003})}\BibitemShut {NoStop}%
\bibitem [{\citenamefont {Tauschinsky}\ \emph {et~al.}(2013)\citenamefont
  {Tauschinsky}, \citenamefont {Newell}, \citenamefont {van~den Heuvell},\ and\
  \citenamefont {Spreeuw}}]{SpreeuwHFS}%
  \BibitemOpen
  \bibfield  {author} {\bibinfo {author} {\bibfnamefont {A.}~\bibnamefont
  {Tauschinsky}}, \bibinfo {author} {\bibfnamefont {R.}~\bibnamefont {Newell}},
  \bibinfo {author} {\bibfnamefont {H.~B. v.~L.}\ \bibnamefont {van~den
  Heuvell}}, \ and\ \bibinfo {author} {\bibfnamefont {R.~J.~C.}\ \bibnamefont
  {Spreeuw}},\ }\href@noop {} {\bibfield  {journal} {\bibinfo  {journal} {Phys.
  Rev. A}\ }\textbf {\bibinfo {volume} {87}},\ \bibinfo {pages} {042522}
  (\bibinfo {year} {2013})}\BibitemShut {NoStop}%
\bibitem [{\citenamefont {Lett}\ \emph {et~al.}(1988)\citenamefont {Lett},
  \citenamefont {Watts}, \citenamefont {Westbrook}, \citenamefont {Phillips},
  \citenamefont {Gould},\ and\ \citenamefont {Metcalf}}]{Metcalf1988}%
  \BibitemOpen
  \bibfield  {author} {\bibinfo {author} {\bibfnamefont {P.~D.}\ \bibnamefont
  {Lett}}, \bibinfo {author} {\bibfnamefont {R.~N.}\ \bibnamefont {Watts}},
  \bibinfo {author} {\bibfnamefont {C.~I.}\ \bibnamefont {Westbrook}}, \bibinfo
  {author} {\bibfnamefont {W.~D.}\ \bibnamefont {Phillips}}, \bibinfo {author}
  {\bibfnamefont {P.~L.}\ \bibnamefont {Gould}}, \ and\ \bibinfo {author}
  {\bibfnamefont {H.~J.}\ \bibnamefont {Metcalf}},\ }\href {\doibase
  10.1103/PhysRevLett.61.169} {\bibfield  {journal} {\bibinfo  {journal} {Phys.
  Rev. Lett.}\ }\textbf {\bibinfo {volume} {61}},\ \bibinfo {pages} {169}
  (\bibinfo {year} {1988})}\BibitemShut {NoStop}%
\bibitem [{\citenamefont {Chu}\ \emph {et~al.}(1985)\citenamefont {Chu},
  \citenamefont {Hollberg}, \citenamefont {Bjorkholm}, \citenamefont {Cable},\
  and\ \citenamefont {Ashkin}}]{Chu1985}%
  \BibitemOpen
  \bibfield  {author} {\bibinfo {author} {\bibfnamefont {S.}~\bibnamefont
  {Chu}}, \bibinfo {author} {\bibfnamefont {L.}~\bibnamefont {Hollberg}},
  \bibinfo {author} {\bibfnamefont {J.~E.}\ \bibnamefont {Bjorkholm}}, \bibinfo
  {author} {\bibfnamefont {A.}~\bibnamefont {Cable}}, \ and\ \bibinfo {author}
  {\bibfnamefont {A.}~\bibnamefont {Ashkin}},\ }\href@noop {} {\bibfield
  {journal} {\bibinfo  {journal} {Phys. Rev. Lett.}\ }\textbf {\bibinfo
  {volume} {55}},\ \bibinfo {pages} {48} (\bibinfo {year} {1985})}\BibitemShut
  {NoStop}%
\bibitem [{\citenamefont {Gallagher}(1994)}]{GallagherBook}%
  \BibitemOpen
  \bibfield  {author} {\bibinfo {author} {\bibfnamefont {T.~F.}\ \bibnamefont
  {Gallagher}},\ }\href@noop {} {\emph {\bibinfo {title} {{R}ydberg Atoms}}}\
  (\bibinfo  {publisher} {Cambridge University Press, Cambridge},\ \bibinfo
  {year} {1994})\BibitemShut {NoStop}%
\bibitem [{\citenamefont {Cisternas}\ \emph {et~al.}(2017)\citenamefont
  {Cisternas}, \citenamefont {de~Hond}, \citenamefont {Lochead}, \citenamefont
  {Spreeuw}, \citenamefont {van~den Heuvell},\ and\ \citenamefont {van
  Druten}}]{CisternasStarkSpec}%
  \BibitemOpen
  \bibfield  {author} {\bibinfo {author} {\bibfnamefont {N.}~\bibnamefont
  {Cisternas}}, \bibinfo {author} {\bibfnamefont {J.}~\bibnamefont {de~Hond}},
  \bibinfo {author} {\bibfnamefont {G.}~\bibnamefont {Lochead}}, \bibinfo
  {author} {\bibfnamefont {R.~J.~C.}\ \bibnamefont {Spreeuw}}, \bibinfo
  {author} {\bibfnamefont {H.~B. v.~L.}\ \bibnamefont {van~den Heuvell}}, \
  and\ \bibinfo {author} {\bibfnamefont {N.~J.}\ \bibnamefont {van Druten}},\
  }\href@noop {} {\bibfield  {journal} {\bibinfo  {journal} {Phys. Rev. A}\
  }\textbf {\bibinfo {volume} {96}},\ \bibinfo {pages} {013425} (\bibinfo
  {year} {2017})}\BibitemShut {NoStop}%
\bibitem [{\citenamefont {Meystre}\ and\ \citenamefont
  {Sargent}(2007)}]{Meystre2007}%
  \BibitemOpen
  \bibfield  {author} {\bibinfo {author} {\bibfnamefont {P.}~\bibnamefont
  {Meystre}}\ and\ \bibinfo {author} {\bibfnamefont {M.}~\bibnamefont
  {Sargent}},\ }\href@noop {} {\emph {\bibinfo {title} {Elements of quantum
  optics}}}\ (\bibinfo  {publisher} {Springer Science \& Business Media},\
  \bibinfo {year} {2007})\BibitemShut {NoStop}%
\bibitem [{\citenamefont {Berman}\ and\ \citenamefont
  {Malinovsky}(2011)}]{BermanBook}%
  \BibitemOpen
  \bibfield  {author} {\bibinfo {author} {\bibfnamefont {P.~R.}\ \bibnamefont
  {Berman}}\ and\ \bibinfo {author} {\bibfnamefont {V.~S.}\ \bibnamefont
  {Malinovsky}},\ }\href@noop {} {\emph {\bibinfo {title} {Principles of Laser
  Spectroscopy and Quantum Optics}}}\ (\bibinfo  {publisher} {Princeton
  University Press},\ \bibinfo {year} {2011})\BibitemShut {NoStop}%
\bibitem [{\citenamefont {Li}\ \emph {et~al.}(2005)\citenamefont {Li},
  \citenamefont {Tanner},\ and\ \citenamefont {Gallagher}}]{GallagherDipole}%
  \BibitemOpen
  \bibfield  {author} {\bibinfo {author} {\bibfnamefont {W.}~\bibnamefont
  {Li}}, \bibinfo {author} {\bibfnamefont {P.~J.}\ \bibnamefont {Tanner}}, \
  and\ \bibinfo {author} {\bibfnamefont {T.~F.}\ \bibnamefont {Gallagher}},\
  }\href@noop {} {\bibfield  {journal} {\bibinfo  {journal} {Phys. Rev. Lett.}\
  }\textbf {\bibinfo {volume} {94}},\ \bibinfo {pages} {173001} (\bibinfo
  {year} {2005})}\BibitemShut {NoStop}%
\bibitem [{\citenamefont {Reinhard}\ \emph {et~al.}(2007)\citenamefont
  {Reinhard}, \citenamefont {Liebisch}, \citenamefont {Knuffman},\ and\
  \citenamefont {Raithel}}]{Reinhard2007}%
  \BibitemOpen
  \bibfield  {author} {\bibinfo {author} {\bibfnamefont {A.}~\bibnamefont
  {Reinhard}}, \bibinfo {author} {\bibfnamefont {T.~C.}\ \bibnamefont
  {Liebisch}}, \bibinfo {author} {\bibfnamefont {B.}~\bibnamefont {Knuffman}},
  \ and\ \bibinfo {author} {\bibfnamefont {G.}~\bibnamefont {Raithel}},\
  }\href@noop {} {\bibfield  {journal} {\bibinfo  {journal} {Phys. Rev. A}\
  }\textbf {\bibinfo {volume} {75}},\ \bibinfo {pages} {032712} (\bibinfo
  {year} {2007})}\BibitemShut {NoStop}%
\bibitem [{\citenamefont {Thaicharoen}\ \emph {et~al.}(2015)\citenamefont
  {Thaicharoen}, \citenamefont {Schwarzkopf},\ and\ \citenamefont
  {Raithel}}]{Pound2015}%
  \BibitemOpen
  \bibfield  {author} {\bibinfo {author} {\bibfnamefont {N.}~\bibnamefont
  {Thaicharoen}}, \bibinfo {author} {\bibfnamefont {A.}~\bibnamefont
  {Schwarzkopf}}, \ and\ \bibinfo {author} {\bibfnamefont {G.}~\bibnamefont
  {Raithel}},\ }\href@noop {} {\bibfield  {journal} {\bibinfo  {journal} {Phys.
  Rev. A}\ }\textbf {\bibinfo {volume} {92}},\ \bibinfo {pages} {040701}
  (\bibinfo {year} {2015})}\BibitemShut {NoStop}%
\bibitem [{\citenamefont {Sa{\ss}mannshausen}\ and\ \citenamefont
  {Deiglmayr}(2016)}]{Sassmannshausen2016}%
  \BibitemOpen
  \bibfield  {author} {\bibinfo {author} {\bibfnamefont {H.}~\bibnamefont
  {Sa{\ss}mannshausen}}\ and\ \bibinfo {author} {\bibfnamefont
  {J.}~\bibnamefont {Deiglmayr}},\ }\href@noop {} {\bibfield  {journal}
  {\bibinfo  {journal} {Phys. Rev. Lett.}\ }\textbf {\bibinfo {volume} {117}},\
  \bibinfo {pages} {083401} (\bibinfo {year} {2016})}\BibitemShut {NoStop}%
\bibitem [{\citenamefont {Boisseau}\ \emph {et~al.}(2002)\citenamefont
  {Boisseau}, \citenamefont {Simbotin},\ and\ \citenamefont
  {C\^ot\'e}}]{Cote2002}%
  \BibitemOpen
  \bibfield  {author} {\bibinfo {author} {\bibfnamefont {C.}~\bibnamefont
  {Boisseau}}, \bibinfo {author} {\bibfnamefont {I.}~\bibnamefont {Simbotin}},
  \ and\ \bibinfo {author} {\bibfnamefont {R.}~\bibnamefont {C\^ot\'e}},\
  }\href {\doibase 10.1103/PhysRevLett.88.133004} {\bibfield  {journal}
  {\bibinfo  {journal} {Phys. Rev. Lett.}\ }\textbf {\bibinfo {volume} {88}},\
  \bibinfo {pages} {133004} (\bibinfo {year} {2002})}\BibitemShut {NoStop}%
\end{thebibliography}%

\end{document}